\documentclass[prd,aps,preprintnumbers,showpacs,nofootinbib,superscriptaddress,notitlepage,floatfix,11pt]{revtex4}
\usepackage{stmaryrd}
\usepackage{epsfig}
\usepackage{amsfonts}
\usepackage{amsmath}
\usepackage{graphicx}
\usepackage{color}
\usepackage{amssymb,amsmath,psfrag,slashed,graphicx}
\usepackage[normalem]{ulem}

\def \nn {\nonumber}

\allowdisplaybreaks

\begin{document}
\title{ Weak decays of doubly heavy baryons: four-body nonleptonic decay channels }

\author{De-Min Li~\footnote{Email: lidm@zzu.edu.cn}}
\affiliation{School of Physics and Microelectronics, Zhengzhou University, Zhengzhou, Henan 450001, China}

\author{Xi-Ruo Zhang}
\affiliation{School of Physics and Microelectronics, Zhengzhou University, Zhengzhou, Henan 450001, China}

\author{Ye Xing~\footnote{Email: xingye\_guang@cumt.edu.cn}}
\affiliation{School of Physics, China University of Mining and Technology, Xuzhou 221000, China}

\author{Ji Xu~\footnote{Email: xuji\_phy@zzu.edu.cn}}
\affiliation{School of Physics and Microelectronics, Zhengzhou University, Zhengzhou, Henan 450001, China}

\begin{abstract}
The LHCb Collaboration announced the observation of doubly charmed baryon through $\Xi_{c c}^{++} \rightarrow \Lambda_{c}^{+} K^{-} \pi^{+} \pi^{+}$ in 2017. Since then, a series of studies of doubly heavy baryons have been presented. $\Xi_{cc}^{++}$ was discovered through nonleptonic four-body decay mode, and experimental data has indicated that the decay modes of $\Xi_{c c}^{++}$ are not saturated by two and three-body intermediate states. In this work, we analyze the four-body weak decays of doubly heavy baryons $\Xi_{cc}^{++}, \Xi_{cc}^+$, and $\Omega_{cc}^+$. Decay amplitudes for various channels are parametrized in terms of SU(3) irreducible amplitudes. We point out that branching fractions for Cabibbo-allowed processes $\Xi_{cc}^{+}\to\Lambda_c^+\pi^+ \pi^0 K^-$, $\Omega_{cc}^{+}\to\Lambda_c^+\pi^+ \overline K^0 K^-$ would be helpful to search for $\Xi_{cc}^+$ and $\Omega_{cc}^+$ in future measurements at experimental facilities like LHC, Belle II, and CEPC.
\end{abstract}

\maketitle

\section{Introduction}
The observation of $\Xi_{cc}^{++}$ by the LHCb Collaboration opens a new field of research studying the nature of baryons containing two heavy quarks, providing a unique platform for testing quantum chromodynamics (QCD)~\cite{Aaij:2017ueg}. The LHCb Collaboration announced the observation of $\Xi_{c c}^{++}$ through $\Xi_{c c}^{++} \rightarrow \Lambda_{c}^{+} K^{-} \pi^{+} \pi^{+}$ channel and confirmed this doubly charmed baryon state by $\Xi_{cc}^{++}\to \Xi_{c}^+ \pi^+$~\cite{Aaij:2018gfl}. Updated measurement of $\Xi_{cc}^{++}$ mass is $3621.55\pm 0.23(\textrm{stat})\pm 0.30(\textrm{syst}) {\rm MeV/c^2}$. With no doubt, this observation is a milestone in hadron physics on both theoretical and experimental sides. Afterwards, more experimental investigations of doubly heavy baryons have been conducted~\cite{Aaij:2018gfl,Aaij:2019dsx,Aaij:2019uaz}. Thus theoretical studies on these baryons  will be of great significance and are urgently requested~\cite{Yu:2017zst,Li:2017ndo,Faessler:2009xn,Albertus:2009ww,Flynn:2007qt,Hernandez:2007qv,Albertus:2006wb,
Ebert:2004ck,Faessler:2001mr,Guo:1998yj,SanchisLozano:1994vh,Wang:2017mqp,Xing:2018bqt,Shi:2019fph}.

In order to understand the properties of doubly heavy baryons, we need to go into the nonperturbative strong interactions inside these particles. For now, Lattice QCD is the only approach that can study nonperturbative strong interactions from first principle. Despite the great progresses in Lattice QCD, hadron structures are still often investigated by approaches such as quark models or QCD sum rules. The quark models~\cite{GellMann:1964nj,G.Zweig:1964} predict the existence of multiplets of baryons and mesons. The lightest four quarks ($u, d, s, c$) can form SU(4) multiplets. In the multiplets, three weakly decaying $qqq$ states with charm quantum number $C=2$ are expected, i.e, one isospin doublet $\Xi_{cc}^{++}(ccu)$, $\Xi_{cc}^{+}(ccd)$ and isospin singlet $\Omega_{cc}^{+}(ccs)$. These states have been searched for a long time~\cite{Ocherashvili:2004hi,Mattson:2002vu,Ratti:2003ez,Aubert:2006qw,Chistov:2006zj,Aaij:2013voa}, finally in 2017 the LHCb Collaboration announced the observation of $\Xi_{cc}^{++}$.

Although this observation marks another significant achievement of quark models, the knowledge of hadron structure is still limited, especially for doubly heavy baryons. The dynamics of weak decays of doubly heavy baryons could be quite different from those of singly heavy baryons due to interference between decay amplitudes of two heavy quarks. Factorization approach has been successfully applied to handle heavy meson decays to separate long and short-distance contributions~\cite{Beneke:1999br,Beneke:2003pa,Keum:2000ph,Keum:2000wi,Lu:2000em,Lu:2000hj,Kurimoto:2001zj}. The short-distance contributions allow the use of perturbation theory and the long-distance contributions are commonly parameterized as low energy inputs such as light-cone distribution amplitudes (LCDAs). As for the issue of doubly heavy baryon decay, neither the short-distance nor the long-distance contributions are available in literature. Various types of weak decays of doubly heavy baryons occur, but unfortunately, a universal factorization approach has not be established yet. This prohibits direct predictions on their decay widths.

\begin{figure}
\includegraphics[width=1\columnwidth]{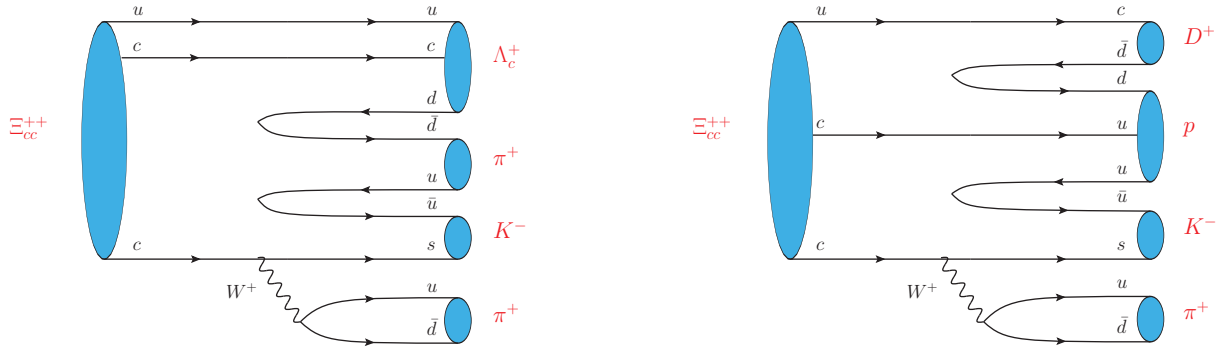}
\caption{Similar Feynman diagrams which contribute to $\Xi_{c c}^{++} \to \Lambda_{c}^{+} K^{-} \pi^{+} \pi^{+}$ (left) and $\Xi_{cc}^{++}\to p D^+ K^- \pi^+$(right).  }
\label{Feynman_diagram_for_four_body}
\end{figure}

Lots of features of doubly charmed baryons are waiting to be discovered from experimental and theoretical sides. Due to approximate isospin symmetry~\cite{Hwang:2008dj,Brodsky:2011zs,Karliner:2017gml}, the masses of the $\Xi_{cc}^{++}$ and $\Xi_{cc}^{+}$ states are expected to differ by only a few $\rm{MeV}/c^2$, but the longer predicted lifetime of $\Xi_{cc}^{++}$ makes it easier to observe than $\Xi_{cc}^{+}$. There is no doubt that searching for $\Xi_{cc}^{+}$ and additional decay modes of $\Xi_{cc}^{++}$ are of prime importance in hadron spectroscopy. The LHCb Collaboration has performed a search for $\Xi_{cc}^{++}$ baryon through the $\Xi_{cc}^{++}\to p D^+ K^- \pi^+$ decay channel, no significant signal is observed~\cite{Aaij:2019dsx}. An upper limit is set on the ratio of branching fractions $\mathcal{R}=\frac{\mathcal{B}\left(\Xi_{c c}^{++} \rightarrow  D^{+} p K^{-} \pi^{+}\right)}{\mathcal{B}\left(\Xi_{c c}^{++} \rightarrow \Lambda_{c}^{+} K^{-} \pi^{+} \pi^{+}\right)}$ with $\mathcal{R}<1.7(2.1) \times 10^{-2}$ at the $90 \%(95 \%)$ confidence level. As shown in Fig.\ref{Feynman_diagram_for_four_body}, the tree-level amplitudes of the exclusive decays of $\Xi_{c c}^{++} \to  D^{+} p K^{-} \pi^{+}$ and $\Xi_{c c}^{++} \rightarrow \Lambda_{c}^{+} K^{-} \pi^{+} \pi^{+}$ are comparable, which implies similar branching fractions of these two modes. A better theoretical understanding of these two decay processes is required to explain the at least two orders of magnitude difference between the branching fractions. Furthermore, there is a long-standing puzzle in the $\Xi_{cc}$ system, observation of the $\Xi_{cc}^{+}$ baryon was reported by the SELEX Collaboration~\cite{Ocherashvili:2004hi,Mattson:2002vu}. However, searches from the FOCUS~\cite{Ratti:2003ez}, BABAR~\cite{Aubert:2006qw}, Belle~\cite{Chistov:2006zj}, and LHCb~\cite{Aaij:2013voa} Collaborations did not find such a state with properties reported by the SELEX Collaboration.

As illustrated before, many features of doubly charmed baryons remain unknown and there is no universal factorization approach established to handle weak decays of doubly charmed baryons. Instead, we will use an optional theoretical tool to analyze doubly charmed baryon decays, the flavor SU(3) symmetry~\cite{Savage:1989ub,Gronau:1994rj,Grinstein:1996us,He:1998rq,Deshpande:2000jp,Deshpande:1994ii,Chiang:2003pm,Chiang:2006ih,
Li:2007bh,Wang:2008rk,Chiang:2008zb,Cheng:2014rfa,He:2014xha,He:2015fwa,He:2015fsa,Lu:2016ogy,Cheng:2016ejf,Cheng:2012xb,Li:2012cfa,Li:2013xsa,
Wang:2017azm,Shi:2017dto,Wang:2018utj}. This approach is an effective way to deal with exclusive decays of hadrons, especially for heavy mesons and heavy baryons. One advantage of the SU(3) analysis is that it is independent of the factorization details, once the branching fractions for a few decay channels have been measured, the flavor SU(3) symmetry can tell us information on the related channels (branching fractions, decay widths, etc.).

$\Xi_{cc}^{++}$ was discovered through nonleptonic four-body decay mode, and experimental data has indicated that the decay modes of $\Xi_{c c}^{++}$ are not saturated by two and three-body intermediate states. This motivates us to study the four-body decays of doubly charmed baryons. Two kinds of weakly decay modes of doubly charmed baryon are considered: $\Xi_{cc}^{++}, \Xi_{cc}^+$ or $\Omega_{cc}^+$ decay into a singly charmed baryon plus three light pseudo-scalar mesons or decay into an octet light baryon with a charmed meson and two light pseudo-scalar mesons. This work is an extension of a series of previous works~\cite{Wang:2017azm,Shi:2017dto,Wang:2018utj}.

The rest of this paper is organized as follows. In Sec.~\ref{sec:particle_multiplet}, we collect the representations for the particle multiplets in the SU(3) symmetry. Sec.~\ref{sec:ccq_nonleptonic} is devoted to the analyzation of four-body nonleptonic decays of doubly charmed baryons. Sec.~\ref{sec:conclusions} consists of a brief summary.

\section{Particle Multiplets}
\label{sec:particle_multiplet}
In this section, we collect the representations for the multiplets of the flavor SU(3) group which are relevant to this paper. The doubly heavy charmed baryons can form an SU(3) triplet which are expressed as:
\begin{eqnarray}
 T_{cc}  = \left(\begin{array}{c}  \Xi^{++}_{cc}(ccu)  \\  \Xi^+_{cc}(ccd)  \\  \Omega^+_{cc}(ccs)
\end{array}\right)\,.
\end{eqnarray}
The singly charmed baryons can form an anti-triplet. In this case, we have the matrix expression:
\begin{eqnarray}
 T_{\bf{c\bar 3}}= \left(\begin{array}{ccc} 0 & \Lambda_c^+  &  \Xi_c^+  \\ -\Lambda_c^+ & 0 & \Xi_c^0 \\ -\Xi_c^+   &  -\Xi_c^0  & 0
  \end{array} \right)\,.
\end{eqnarray}
The light baryons can form an SU(3) octet which has the expression:
\begin{eqnarray}
T_8= \left(\begin{array}{ccc} \frac{1}{\sqrt{2}}\Sigma^0+\frac{1}{\sqrt{6}}\Lambda^0 & \Sigma^+  &  p  \\ \Sigma^-  &  -\frac{1}{\sqrt{2}}\Sigma^0+\frac{1}{\sqrt{6}}\Lambda^0 & n \\ \Xi^-   & \Xi^0  & -\sqrt{\frac{2}{3}}\Lambda^0
  \end{array} \right) \,.
\end{eqnarray}
In the meson sector, the light pseudo-scalar mesons form an octet:
\begin{eqnarray}
 M_{8}=\begin{pmatrix}
 \frac{\pi^0}{\sqrt{2}}+\frac{\eta}{\sqrt{6}}  &\pi^+ & K^+\\
 \pi^-&-\frac{\pi^0}{\sqrt{2}}+\frac{\eta}{\sqrt{6}}&{K^0}\\
 K^-&\bar K^0 &-2\frac{\eta}{\sqrt{6}}
 \end{pmatrix} \,.
\end{eqnarray}
The charmed mesons form an SU(3) anti-triplet:
\begin{eqnarray}
D_i=\left(\begin{array}{ccc} D^0, & D^+, & D^+_s  \end{array} \right) \,.
\end{eqnarray}

\section{Non-Leptonic $\Xi_{cc}$ and $\Omega_{cc}$ decays}
\label{sec:ccq_nonleptonic}
Considering Cabibbo angle, nonleptonic charm-quark decays into light quarks are categorized into three groups: Cabibbo allowed, singly Cabibbo
suppressed, and doubly Cabibbo suppressed,
\begin{eqnarray}
 c\to s \bar d u,  \;\;\; c\to u \bar dd/\bar ss, \;\;\; c\to  d \bar s u.
\end{eqnarray}
These tree operators transform under the flavor SU(3) symmetry as ${\bf  3}\otimes {\bf\bar 3}\otimes {\bf3}={\bf  3}\oplus {\bf  3}\oplus {\bf\bar 6}\oplus {\bf {15}}$. Thus the effective Hamiltonian can be decomposed in terms of a vector $H_3$; a traceless tensor antisymmetric in upper indices, $H_{\bf\overline6}$; a traceless tensor symmetric in upper indices, $H_{\bf {15}}$. For charm quark decays, the representation $H_3$ will vanish as an approximation by taking $V_{cd}^*V_{ud} = -V_{cs}^*V_{us}\simeq -\sin(\theta_C)$~\cite{Wang:2017azm}. The nonzero components of hadron-level Hamiltonian are listed below:
\begin{eqnarray}
(H_{\overline 6})^{31}_2=-(H_{\overline 6})^{13}_2=1 \,,\;\;\;
 (H_{15})^{31}_2= (H_{15})^{13}_2=1 \,, \qquad&&\rm{Cabibbo~allowed \,,} \nn\\
 (H_{\overline 6})^{31}_3 =-(H_{\overline 6})^{13}_3 =(H_{\overline 6})^{12}_2 =-(H_{\overline 6})^{21}_2 =\sin(\theta_C)\,, \qquad&&\rm{Singly~Cabibbo~suppressed \,,} \nn\\
 (H_{15})^{31}_3= (H_{15})^{13}_3=-(H_{15})^{12}_2=-(H_{15})^{21}_2= \sin(\theta_C)\,,  \qquad&&\rm{Singly~Cabibbo~suppressed \,,} \nn\\
 (H_{\overline 6})^{21}_3=-(H_{\overline 6})^{12}_3=\sin^2\theta_C,\;\;
 (H_{15})^{21}_3= (H_{15})^{12}_3=\sin^2\theta_C \,, \qquad&&\rm{Doubly~Cabibbo~suppressed \,.} \nn\\
\end{eqnarray}

\subsection{Decays into a singly charmed baryon and three light pseudo-scalar mesons}
With the above expressions, the effective Hamiltonian for $\Xi_{cc}$ and $\Omega_{cc}$ decay into an antitriplet charmed baryon and three light pseudo-scalar mesons can be derived as
\begin{eqnarray}\label{H:1}
 {\cal H}_{\textit{eff}}&=& a_1  (T_{cc})^i  (\overline T_{c\bar 3})_{[ij]}  M^{j}_{k}  M^{k}_{l}  M^{n}_{m} (H_{\overline6})^{lm}_{n}
                           +a_2  (T_{cc})^i  (\overline T_{c\bar 3})_{[ij]}  M^{j}_{k}  M^{m}_{n}  M^{n}_{l} (H_{\overline6})^{kl}_{m} \nn\\
                         &&+a_3  (T_{cc})^i  (\overline T_{c\bar 3})_{[ij]}  M^{n}_{m}  M^{m}_{n}  M^{l}_{k} (H_{\overline6})^{jk}_{l}
                           +a_4  (T_{cc})^i  (\overline T_{c\bar 3})_{[ij]}  M^{n}_{m}  M^{l}_{n}  M^{m}_{k} (H_{\overline6})^{jk}_{l} \nn\\
                         &&+a_5  (T_{cc})^i  (\overline T_{c\bar 3})_{[jk]}  M^{j}_{i}  M^{k}_{l}  M^{n}_{m} (H_{\overline6})^{lm}_{n}
                           +a_6  (T_{cc})^i  (\overline T_{c\bar 3})_{[jk]}  M^{j}_{i}  M^{m}_{l}  M^{n}_{m} (H_{\overline6})^{kl}_{n} \nn\\
                         &&+a_7  (T_{cc})^i  (\overline T_{c\bar 3})_{[kl]}  M^{j}_{i}  M^{k}_{j}  M^{n}_{m} (H_{\overline6})^{lm}_{n}
                           +a_8  (T_{cc})^i  (\overline T_{c\bar 3})_{[mn]}  M^{j}_{i}  M^{k}_{j}  M^{l}_{k} (H_{\overline6})^{mn}_{l} \nn\\
                         &&+a_9  (T_{cc})^i  (\overline T_{c\bar 3})_{[ln]}  M^{j}_{i}  M^{k}_{j}  M^{n}_{m} (H_{\overline6})^{lm}_{k}
                           +a_{10}  (T_{cc})^i  (\overline T_{c\bar 3})_{[kl]}  M^{j}_{i}  M^{m}_{n}  M^{n}_{m} (H_{\overline6})^{kl}_{j} \nn\\
                         &&+a_{11}  (T_{cc})^i  (\overline T_{c\bar 3})_{[km]}  M^{j}_{i}  M^{n}_{l}  M^{m}_{n} (H_{\overline6})^{kl}_{j}
                           +a_{12}  (T_{cc})^i  (\overline T_{c\bar 3})_{[mn]}  M^{j}_{i}  M^{m}_{k}  M^{n}_{l} (H_{\overline6})^{kl}_{j} \nn\\
                         &&+a_{13}  (T_{cc})^i  (\overline T_{c\bar 3})_{[jk]}  M^{l}_{m}  M^{n}_{l}  M^{m}_{n} (H_{\overline6})^{jk}_{i}
                           +a_{14}  (T_{cc})^i  (\overline T_{c\bar 3})_{[jl]}  M^{l}_{k}  M^{m}_{n}  M^{n}_{m} (H_{\overline6})^{jk}_{i} \nn\\
                         &&+a_{15}  (T_{cc})^i  (\overline T_{c\bar 3})_{[jl]}  M^{n}_{k}  M^{l}_{m}  M^{m}_{n} (H_{\overline6})^{jk}_{i}
                           +a_{16}  (T_{cc})^i  (\overline T_{c\bar 3})_{[lm]}  M^{l}_{j}  M^{n}_{k}  M^{m}_{n} (H_{\overline6})^{jk}_{i} \nn\\
                         &&+\Big( (H_{\overline6}) \to (H_{15}), a\to b\Big) \,.
\end{eqnarray}
The last term means that there are implicit terms corresponding to $(H_{15})$ by simply replacing $(H_{\overline6}) \to (H_{15})$ and $a\to b$. Here the $a_i$ and $b_i$ are SU(3) irreducible nonperturbative amplitudes. Due to anti-symmetric properties of indices $i,j$ in $(\overline T_{c\bar 3})_{[ij]}$ and symmetric properties of $i,j$ in $(H_{15})^{ij}_{k}$, one can come to the conclusion that $b_8, b_{10}, b_{12}$, and $b_{13}$ are zero. Besides, we have checked that some matrix elements in the Hamiltonian are not independent, for example, $a_1$ and $a_2$ are always presented as $a_1-a_2$ in the amplitude. In addition, there are other coefficients that appear in groups, such as $a_3$ with $a_4$, $a_5$ with $a_6$, $b_3$ with $b_4$, and $b_{14}$ with $b_{15}$. So we can remove $a_2, a_4, a_6, b_4, b_{15}$ terms in the expanded amplitude.

\begin{figure}
\includegraphics[width=1\columnwidth]{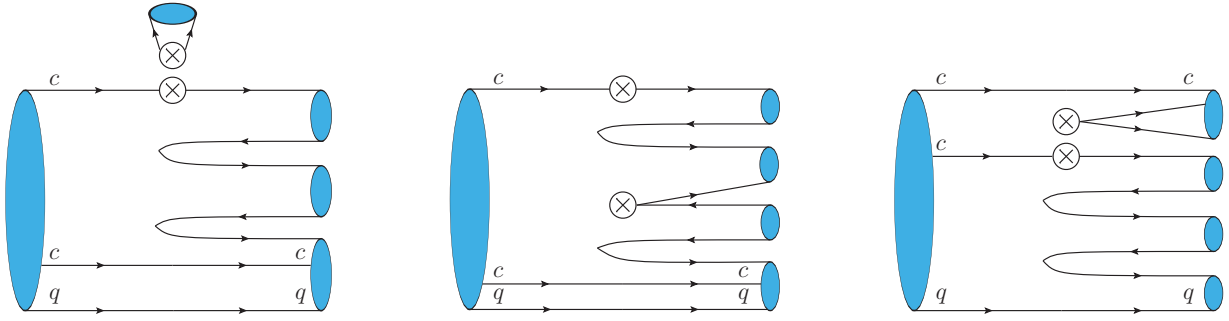}
\caption{ Feynman diagrams for $\Xi_{cc}$ and $\Omega_{cc}$ decays into a singly charmed baryon and three light pseudo-scalar mesons. }
\label{feynman1}
\end{figure}

Feynman diagrams for some channels are presented in Fig.~\ref{feynman1} for illustration. The decay amplitudes for different channels can be deduced by expanding the Hamiltonian in Eq.(\ref{H:1}). These channels are collected in three tables according to their dependence on $\sin(\theta_C)$. Table~\ref{tab:First_type_allowed}, Table~\ref{tab:First_type_singly}, and Table~\ref{tab:First_type_doubly} represent the Cabibbo-allowed, singly Cabibbo-suppressed, and doubly Cabibbo-suppressed processes, respectively.

\begin{table}
\begin{tiny}
  		\caption{Cabibbo-allowed channels for doubly charmed baryon decays into an antitriplet charmed baryon and three light mesons.}\label{tab:First_type_allowed}
\begin{tabular}{|c|c|c|c|c|c|c|c}\hline\hline
  			channel & amplitude \\\hline
  	        $\Xi_{cc}^{++}\to \Lambda_c^+  \pi^+   \pi^+   K^-  $ & $ 2\left(a_1-a_5-a_{11}+2 a_{12}+b_1+b_2-b_5-b_6+b_{11}\right)$\\\hline
  			$\Xi_{cc}^{++}\to \Lambda_c^+  \pi^+   \pi^0   \overline K^0  $ & $ \frac{-a_1+a_5+a_{11}-2 a_{12}-b_1-b_2+b_5+b_6-b_{11}}{\sqrt{2}}$\\\hline
  			$\Xi_{cc}^{++}\to \Lambda_c^+  \pi^+   \overline K^0   \eta  $ & $ \frac{-3 a_1+a_5+2 a_7-2 a_9+a_{11}-2 a_{12}+b_1+b_2+b_5+b_6-2 b_7+2 b_9-b_{11}}{\sqrt{6}}$\\\hline
  			$\Xi_{cc}^{++}\to \Xi_c^+  \pi^+   \pi^+   \pi^-  $ & $ 2\left(2 a_3+a_7-2 a_8-4 a_{10}-a_{11}+2 b_3+b_7-b_{11}\right)$\\\hline
  			$\Xi_{cc}^{++}\to \Xi_c^+  \pi^+   \pi^0   \pi^0  $ & $ 2 a_3+a_7-2 a_8-4 a_{10}-a_{11}+2 b_3+b_7-b_{11}$\\\hline
  			$\Xi_{cc}^{++}\to \Xi_c^+  \pi^+   \pi^0   \eta  $ & $ -\frac{a_5-a_7+a_9+a_{11}-2 a_{12}+b_5-b_6-b_7+b_9+b_{11}}{\sqrt{3}}$\\\hline
  			$\Xi_{cc}^{++}\to \Xi_c^+  \pi^+   K^+   K^-  $ & $ a_1+2 a_3-a_5+a_7-2 a_8-4 a_{10}-2 a_{11}+2 a_{12}+b_1+b_2+2 b_3-b_5-b_6+b_7$\\\hline
  			$\Xi_{cc}^{++}\to \Xi_c^+  \pi^+   K^0   \overline K^0  $ & $ 2 a_3+a_7-2 a_8-4 a_{10}-a_{11}+2 b_1+2 b_3-b_7+b_{11}$\\\hline
  			$\Xi_{cc}^{++}\to \Xi_c^+  \pi^0   K^+   \overline K^0  $ & $ -\frac{a_1-a_7+a_9-b_1+b_2-2 b_6+b_7+b_9}{\sqrt{2}}$\\\hline
  			$\Xi_{cc}^{++}\to \Xi_c^0  \pi^+   \pi^+   \eta  $ & $ -\frac{2\sqrt{2}}{\sqrt{3}} \left(a_5-a_7+a_9+a_{11}-2 a_{12}+b_5-b_6-b_7+b_9+b_{11}\right)$\\\hline
  			$\Xi_{cc}^{++}\to \Xi_c^0  \pi^+   K^+   \overline K^0  $ & $ -a_5+a_7-a_9-a_{11}+2 a_{12}-b_5+b_6+b_7-b_9-b_{11}$\\\hline
  			$\Xi_{cc}^{+}\to \Lambda_c^+  \pi^+   \pi^0   K^-  $ & $ \frac{-a_1+a_5+a_{11}-2 a_{12}-b_1-b_2+b_5+b_6-b_{11}}{\sqrt{2}}$\\\hline
  			$\Xi_{cc}^{+}\to \Lambda_c^+  \pi^+   \pi^-   \overline K^0  $ & $ 2 a_3+a_7-a_9-2 a_{14}-a_{15}-2 b_1-2 b_3+2 b_5-b_7+b_9+2 b_{14}-2 b_{16}$\\\hline
  			$\Xi_{cc}^{+}\to \Lambda_c^+  \pi^+   K^-   \eta  $ & $ \frac{a_1-a_5-a_{11}+2 a_{12}+b_1-3 b_2-b_5-b_6+b_{11}+4 b_{16}}{\sqrt{6}}$\\\hline
  			$\Xi_{cc}^{+}\to \Lambda_c^+  \pi^0   \overline K^0   \eta  $ & $ \frac{a_1-a_7+a_9-b_1+b_2+b_7-b_9-2 b_{16}}{\sqrt{3}}$\\\hline
  			$\Xi_{cc}^{+}\to \Xi_c^+  \pi^+   \pi^-   \eta  $ & $ -\frac{2 a_5-2 a_7+6 a_8-a_9+4 a_{10}+a_{11}+12 a_{13}-2 a_{14}+3 a_{15}-2 a_{16}+2 b_5-2 b_6-2 b_7+3 b_9+b_{11}+6 b_{14}-2 b_{16}}{\sqrt{6}}$\\\hline
  			$\Xi_{cc}^{+}\to \Xi_c^+  \pi^+   K^0   K^-  $ & $ -a_5+a_7-2 a_8-6 a_{13}-2 a_{15}+a_{16}-b_5-b_6+b_7+b_{16}$\\\hline
  			$\Xi_{cc}^{+}\to \Xi_c^+  \pi^0   \pi^0   \pi^0  $ & $ \frac{3\left(2 a_8-a_9+4 a_{10}+a_{11}-2 a_{14}-a_{15}-b_9+b_{11}-2 b_{14}\right)}{ \sqrt{2}}$\\\hline
  			$\Xi_{cc}^{+}\to \Xi_c^+  \pi^0   \pi^0   \eta  $ & $ \frac{-6 a_8+3 a_9-4 a_{10}+a_{11}-4 a_{12}-12 a_{13}+2 a_{14}-3 a_{15}+2 a_{16}-b_9+b_{11}-6 b_{14}+2 b_{16}}{ \sqrt{6}}$\\\hline
  			$\Xi_{cc}^{+}\to \Xi_c^+  \pi^0   K^+   K^-  $ & $ \frac{4 a_{10}+2 a_{11}-2 a_{12}-6 a_{13}-2 a_{14}-3 a_{15}+a_{16}-2 b_{14}+b_{16}}{\sqrt{2}}$\\\hline
  			$\Xi_{cc}^{+}\to \Xi_c^+  \pi^0   K^0   \overline K^0  $ & $ \frac{a_5-a_7+4 a_8-a_9+4 a_{10}+a_{11}+6 a_{13}-2 a_{14}+a_{15}-a_{16}-b_5+b_6+b_7-b_9-b_{11}-2 b_{14}+b_{16}}{\sqrt{2}}$\\\hline
  			$\Xi_{cc}^{+}\to \Xi_c^+  \pi^-   K^+   \overline K^0  $ & $ -a_5+a_7-2 a_8-6 a_{13}-2 a_{15}+a_{16}+b_5+b_6-b_7-b_{16}$\\\hline
  			$\Xi_{cc}^{+}\to \Xi_c^0  \pi^+   \pi^+   \pi^-  $ & $ 2\left(2 a_3+a_7-a_9-2 a_{14}-a_{15}+2 b_3+b_7-b_9-2 b_{14}\right)$\\\hline
  			$\Xi_{cc}^{+}\to \Xi_c^0  \pi^+   \pi^0   \pi^0  $ & $ 2 a_3+a_7-a_9-2 a_{14}-a_{15}+2 b_3+b_7-b_9-2 b_{14}$\\\hline
  			$\Xi_{cc}^{+}\to \Xi_c^0  \pi^+   \pi^0   \eta  $ & $ \frac{a_5-a_7+a_9+a_{11}-2 a_{12}+b_5-b_6-b_7+b_9+b_{11}}{\sqrt{3}}$\\\hline
  			$\Xi_{cc}^{+}\to \Xi_c^0  \pi^+   K^+   K^-  $ & $ a_1+2 a_3-2 a_{14}-a_{15}+b_1+b_2+2 b_3-2 b_{14}$\\\hline
  			$\Xi_{cc}^{+}\to \Xi_c^0  \pi^+   K^0   \overline K^0  $ & $ 2 a_3+a_7-a_9-2 a_{14}-a_{15}+2 b_1+2 b_3-2 b_5+b_7-b_9-2 b_{14}+2 b_{16}$\\\hline
  			$\Xi_{cc}^{+}\to \Xi_c^0  \pi^0   K^+   \overline K^0  $ & $ \frac{-a_1+a_5+a_{11}-2 a_{12}+b_1-b_2-b_5-b_6+b_{11}+2 b_{16}}{\sqrt{2}}$\\\hline
  			$\Omega_{cc}^{+}\to \Lambda_c^+  \pi^+   \overline K^0   K^-  $ & $ -a_5+a_7-a_9-a_{11}+2 a_{12}+b_5-b_6-b_7+b_9+b_{11}$\\\hline
  			$\Omega_{cc}^{+}\to \Lambda_c^+  \pi^0   \overline K^0   \overline K^0  $ & $ \sqrt{2}\left(a_5-a_7+a_9+a_{11}-2 a_{12}-b_5+b_6+b_7-b_9-b_{11}\right)$\\\hline
  			$\Omega_{cc}^{+}\to \Lambda_c^+  \overline K^0   \overline K^0   \eta  $ & $\sqrt{\frac{2}{3}} \left(a_5-a_7+a_9+a_{11}-2 a_{12}-b_5+b_6+b_7-b_9-b_{11}\right)$\\\hline
  			$\Omega_{cc}^{+}\to \Xi_c^+  \pi^+   \pi^0   K^-  $ & $ -\frac{a_1-a_7+a_9+b_1+b_2-b_7+b_9}{\sqrt{2}}$\\\hline
  			$\Omega_{cc}^{+}\to \Xi_c^+  \pi^+   \pi^-   \overline K^0  $ & $ 2 a_3+a_7-2 a_8-4 a_{10}-a_{11}-2 b_1-2 b_3+b_7-b_{11}$\\\hline
  			$\Omega_{cc}^{+}\to \Xi_c^+  \pi^+   K^-   \eta  $ & $ \frac{a_1-a_7+a_9+b_1-3 b_2+4 b_6-b_7-3 b_9}{\sqrt{6}}$\\\hline
  			$\Omega_{cc}^{+}\to \Xi_c^+  \pi^0   \pi^0   \overline K^0  $ & $ a_1+2 a_3-2 a_8+a_9-4 a_{10}-a_{11}-b_1+b_2-2 b_3+b_9-b_{11}$\\\hline
  			$\Omega_{cc}^{+}\to \Xi_c^+  \pi^0   \overline K^0   \eta  $ & $ \frac{a_1-a_5-a_{11}+2 a_{12}-b_1+b_2+b_5-b_6+2 b_9-b_{11}}{\sqrt{3}}$\\\hline
  			$\Omega_{cc}^{+}\to \Xi_c^0  \pi^+   \pi^+   K^-  $ & $ 2\left(-a_1+a_7-a_9-b_1-b_2+b_7-b_9\right)$\\\hline
  			$\Omega_{cc}^{+}\to \Xi_c^0  \pi^+   \pi^0   \overline K^0  $ & $ \frac{a_1-a_7+a_9+b_1+b_2-b_7+b_9}{\sqrt{2}}$\\\hline
  			$\Omega_{cc}^{+}\to \Xi_c^0  \pi^+   \overline K^0   \eta  $ & $ \frac{3 a_1-2 a_5-a_7+a_9-2 a_{11}+4 a_{12}-b_1-b_2+2 b_5+2 b_6-b_7+b_9-2 b_{11}}{\sqrt{6}}$\\\hline
  			\hline
\end{tabular}
\end{tiny}
\end{table}

\begin{table}
  	\begin{tiny}
  		\caption{Singly Cabibbo-suppressed channels for doubly charmed baryon decays into an antitriplet charmed baryon and three light mesons.}\label{tab:First_type_singly}
  \begin{tabular}{|c|c|c|c|c|c|c|c}\hline\hline
  			channel & amplitude \\\hline
  			$\Xi_{cc}^{++}\to \Lambda_c^+  \pi^+   \pi^+   \pi^-  $ & $ -2\sin(\theta_C) \left(a_1+2 a_3-a_5+a_7-2 a_8-4 a_{10}-2 a_{11}+2 a_{12}+b_1+b_2+2 b_3-b_5-b_6+b_7\right)$\\\hline
  			$\Xi_{cc}^{++}\to \Lambda_c^+  \pi^+   \pi^0   \pi^0  $ & $ -\sin(\theta_C) \left(a_1+2 a_3-a_5+a_7-2 a_8-4 a_{10}-2 a_{11}+2 a_{12}+b_1+b_2+2 b_3-b_5-b_6+b_7\right)$\\\hline
  			$\Xi_{cc}^{++}\to \Lambda_c^+  \pi^+   \pi^0   \eta  $ & $ \frac{2 \left(b_1+b_2-b_6-b_7+b_9+b_{11}\right) \sin(\theta_C)}{\sqrt{3}}$\\\hline
  			$\Xi_{cc}^{++}\to \Lambda_c^+  \pi^+   K^+   K^-  $ & $ \sin(\theta_C) \left(a_1-2 a_3-a_5-a_7+2 a_8+4 a_{10}+2 a_{12}+b_1+b_2-2 b_3-b_5-b_6-b_7+2 b_{11}\right)$\\\hline
  			$\Xi_{cc}^{++}\to \Lambda_c^+  \pi^+   K^0   \overline K^0  $ & $ -\sin(\theta_C) \left(a_1+2 a_3-2 a_8+a_9-4 a_{10}-a_{11}+b_1-b_2+2 b_3-b_9+b_{11}\right)$\\\hline
  			$\Xi_{cc}^{++}\to \Lambda_c^+  \pi^+   \eta   \eta  $ & $ \frac{1}{3} \sin(\theta_C) \left(5 a_1-6 a_3-a_5-7 a_7+6 a_8+4 a_9+12 a_{10}+2 a_{11}+2 a_{12}-7 b_1+b_2-6 b_3-b_5-5 b_6+5 b_7\right)$\\\hline
  			$\Xi_{cc}^{++}\to \Lambda_c^+  \pi^0   K^+   \overline K^0  $ & $ -\frac{\sin(\theta_C) \left(2 a_1-a_5-a_7+a_9-a_{11}+2 a_{12}-2 b_2-b_5+b_6+b_7-3 b_9+b_{11}\right)}{\sqrt{2}}$\\\hline
  			$\Xi_{cc}^{++}\to \Xi_c^+  \pi^+   \pi^0   K^0  $ & $ -\frac{\sin(\theta_C) \left(a_5-a_7+a_9+a_{11}-2 a_{12}-2 b_1+b_5-b_6+b_7+b_9-b_{11}\right)}{\sqrt{2}}$\\\hline
  			$\Xi_{cc}^{++}\to \Xi_c^+  \pi^+   \pi^-   K^+  $ & $ \sin(\theta_C) \left(-a_1+2 a_3+a_5+a_7-2 a_8-4 a_{10}-2 a_{12}-b_1-b_2+2 b_3+b_5+b_6+b_7-2 b_{11}\right)$\\\hline
  			$\Xi_{cc}^{++}\to \Xi_c^+  \pi^0   \pi^0   K^+  $ & $ -\sin(\theta_C) \left(a_1-2 a_3-2 a_7+2 a_8+a_9+4 a_{10}+a_{11}-b_1+b_2-2 b_3-2 b_6+b_9+b_{11}\right)$\\\hline
  			$\Xi_{cc}^{++}\to \Xi_c^0  \pi^+   \pi^+   K^0  $ & $ -2\sin(\theta_C) \left(a_5-a_7+a_9+a_{11}-2 a_{12}+b_5-b_6-b_7+b_9+b_{11}\right)$\\\hline
  			$\Xi_{cc}^{++}\to \Xi_c^0  \pi^+   \pi^0   K^+  $ & $ -\frac{\sin(\theta_C) \left(a_5-a_7+a_9+a_{11}-2 a_{12}+b_5-b_6-b_7+b_9+b_{11}\right)}{\sqrt{2}}$\\\hline
  			$\Xi_{cc}^{+}\to \Lambda_c^+  \pi^+   \pi^0   \pi^-  $ & $ \frac{\sin(\theta_C) \left(a_1+2 a_3-a_5+a_7-2 a_8-4 a_{10}-2 a_{11}+2 a_{12}-b_1+b_2-2 b_3+b_5-b_6-b_7+2 b_9+4 b_{14}-2 b_{16}\right)}{\sqrt{2}}$\\\hline
  			$\Xi_{cc}^{+}\to \Lambda_c^+  \pi^+   \pi^-   \eta  $ & $ -\frac{\sin(\theta_C) \left(a_1+6 a_3-3 a_5+5 a_7-6 a_8-2 a_9-4 a_{10}-2 a_{11}+2 a_{12}-12 a_{13}-4 a_{14}-6 a_{15}+2 a_{16}-5 b_1-3 b_2-6 b_3+3 b_5+b_6-b_7\right)}{\sqrt{6}}$\\\hline
  			$\Xi_{cc}^{+}\to \Lambda_c^+  \pi^+   K^0   K^-  $ & $ \sin(\theta_C) \left(a_1-a_7+2 a_8-a_{11}+2 a_{12}+6 a_{13}+2 a_{15}-a_{16}+b_1-b_2-b_7+b_{11}+b_{16}\right)$\\\hline
  			$\Xi_{cc}^{+}\to \Lambda_c^+  \pi^0   \pi^0   \pi^0  $ & $ \frac{3\sin(\theta_C) \left(a_1+2 a_3-a_5+a_7-2 a_8-4 a_{10}-2 a_{11}+2 a_{12}-b_1+b_2-2 b_3+b_5-b_6-b_7+2 b_9+4 b_{14}-2 b_{16}\right)}{\sqrt{2}}$\\\hline
  			$\Xi_{cc}^{+}\to \Lambda_c^+  \pi^0   K^+   K^-  $ & $ \frac{\sin(\theta_C) \left(2 a_3+a_5-4 a_{10}-a_{11}+6 a_{13}+2 a_{15}-a_{16}-2 b_1-2 b_2-2 b_3+b_5+b_6-b_{11}+4 b_{14}-b_{16}\right)}{\sqrt{2}}$\\\hline
  			$\Xi_{cc}^{+}\to \Lambda_c^+  \pi^0   K^0   \overline K^0  $ & $ \frac{\sin(\theta_C) \left(2 a_3+a_7-4 a_8+a_9-4 a_{10}-2 a_{12}-6 a_{13}-2 a_{15}+a_{16}-2 b_3-b_7+b_9+4 b_{14}-b_{16}\right)}{\sqrt{2}}$\\\hline
  			$\Xi_{cc}^{+}\to \Lambda_c^+  \pi^0   \eta   \eta  $ & $ -\frac{\sin(\theta_C) \left(5 a_1-6 a_3-a_5-7 a_7+6 a_8+4 a_9+12 a_{10}+2 a_{11}+2 a_{12}-5 b_1+5 b_2+6 b_3+b_5-b_6+7 b_7-2 b_9-4 b_{11}-12 b_{14}-2 b_{16}\right)}{3 \sqrt{2}}$\\\hline
  			$\Xi_{cc}^{+}\to \Lambda_c^+  \pi^-   K^+   \overline K^0  $ & $ -\sin(\theta_C) \left(a_1-a_5-2 a_8+a_9-6 a_{13}-2 a_{15}+a_{16}+b_1-b_2-b_5+b_6-b_9+b_{16}\right)$\\\hline
  			$\Xi_{cc}^{+}\to \Xi_c^+  \pi^+   \pi^-   K^0  $ & $ -\sin(\theta_C) \left(2 a_8-a_9+4 a_{10}+a_{11}-2 a_{14}-a_{15}-2 b_6+b_9+b_{11}+2 b_{14}\right)$\\\hline
  			$\Xi_{cc}^{+}\to \Xi_c^+  \pi^0   \pi^0   K^0  $ & $  \sin(\theta_C) \left(a_5-a_7-2 a_8+2 a_9-4 a_{10}-2 a_{12}+2 a_{14}+a_{15}-b_5+b_6+b_7-2 b_{11}-2 b_{14}+2 b_{16}\right)$\\\hline
  			$\Xi_{cc}^{+}\to \Xi_c^+  \pi^0   \pi^-   K^+  $ & $ -\frac{\sin(\theta_C) \left(a_5-a_7+a_9+a_{11}-2 a_{12}-b_5-b_6+b_7+b_9-b_{11}+2 b_{16}\right)}{\sqrt{2}}$\\\hline
  			$\Xi_{cc}^{+}\to \Xi_c^+  \pi^0   K^0   \eta  $ & $ -\frac{\sin(\theta_C) \left(a_5-a_7+a_9+a_{11}-2 a_{12}-b_5+b_6+b_7-3 b_9+b_{11}-2 b_{16}\right)}{\sqrt{3}}$\\\hline
  			$\Xi_{cc}^{+}\to \Xi_c^+  \pi^-   K^+   \eta  $ & $ \frac{\sin(\theta_C) \left(a_5-a_7+a_9+a_{11}-2 a_{12}-5 b_5-b_6+5 b_7-3 b_9-b_{11}+2 b_{16}\right)}{\sqrt{6}}$\\\hline
  			$\Xi_{cc}^{+}\to \Xi_c^0  \pi^+   \pi^0   K^0  $ & $ \frac{\sin(\theta_C) \left(a_5-a_7+a_9+a_{11}-2 a_{12}+2 b_1-b_5-b_6-b_7+b_9+b_{11}+2 b_{16}\right)}{\sqrt{2}}$\\\hline
  			$\Xi_{cc}^{+}\to \Xi_c^0  \pi^+   \pi^-   K^+  $ & $ -\sin(\theta_C) \left(a_1-2 a_3-2 a_7+2 a_9+2 a_{14}+a_{15}+b_1+b_2-2 b_3-2 b_7+2 b_9+2 b_{14}\right)$\\\hline
  			$\Xi_{cc}^{+}\to \Xi_c^0  \pi^+   K^0   \eta  $ & $ \frac{\sin(\theta_C) \left(4 a_1-3 a_5-a_7+a_9-3 a_{11}+6 a_{12}-2 b_1-4 b_2+3 b_5+3 b_6-b_7+b_9-3 b_{11}+2 b_{16}\right)}{\sqrt{6}}$\\\hline
  			$\Xi_{cc}^{+}\to \Xi_c^0  \pi^0   K^+   \eta  $ & $ \frac{\sin(\theta_C) \left(2 a_1-a_5-a_7+a_9-a_{11}+2 a_{12}-2 b_1+3 b_5+b_6-b_7+b_9-b_{11}-2 b_{16}\right)}{\sqrt{3}}$\\\hline
  			$\Omega_{cc}^{+}\to \Lambda_c^+  \pi^+   \pi^0   K^-  $ & $ \sqrt{2} \left(b_5-b_7+b_9\right) \sin(\theta_C)$\\\hline
  			$\Omega_{cc}^{+}\to \Lambda_c^+  \pi^+   \pi^-   \overline K^0  $ & $ \sin(\theta_C) \left(a_5-a_7+2 a_8+4 a_{10}+2 a_{11}-2 a_{12}-2 a_{14}-a_{15}+b_5+b_6-b_7+2 b_{14}-2 b_{16}\right)$\\\hline
  			$\Omega_{cc}^{+}\to \Lambda_c^+  \pi^+   K^-   \eta  $ & $ \sqrt{\frac{2}{3}} \sin(\theta_C) \left(a_5-a_7+a_9+a_{11}-2 a_{12}-2 b_5-b_6+2 b_7-b_{11}+2 b_{16}\right)$\\\hline
  			$\Omega_{cc}^{+}\to \Lambda_c^+  \pi^0   \pi^0   \overline K^0  $ & $ \sin(\theta_C) \left(a_5-a_7+2 a_8+4 a_{10}+2 a_{11}-2 a_{12}-2 a_{14}-a_{15}-b_5+b_6+b_7-2 b_9+2 b_{14}-2 b_{16}\right)$\\\hline
  			$\Omega_{cc}^{+}\to \Lambda_c^+  \pi^0   \overline K^0   \eta  $ & $ -\frac{\sin(\theta_C) \left(a_5-a_7+a_9+a_{11}-2 a_{12}-b_5+b_6+b_7+b_9-3 b_{11}+2 b_{16}\right)}{\sqrt{3}}$\\\hline
  			$\Omega_{cc}^{+}\to \Lambda_c^+  K^+   \overline K^0   K^-  $ & $ \sin(\theta_C) \left(2 a_8-a_9+4 a_{10}+a_{11}-2 a_{14}-a_{15}-2 b_6+b_9+b_{11}+2 b_{14}\right)$\\\hline
  			$\Omega_{cc}^{+}\to \Lambda_c^+  K^0   \overline K^0   \overline K^0  $ & $ 2\sin(\theta_C) \left(2 a_8-a_9+4 a_{10}+a_{11}-2 a_{14}-a_{15}+b_9-b_{11}+2 b_{14}\right)$\\\hline
  			$\Omega_{cc}^{+}\to \Xi_c^+  \pi^+   \pi^0   \pi^-  $ & $ \frac{\sin(\theta_C) \left(a_1+2 a_3-2 a_{14}-a_{15}-b_1+b_2-2 b_3-2 b_{14}\right)}{\sqrt{2}}$\\\hline
  			$\Omega_{cc}^{+}\to \Xi_c^+  \pi^+   \pi^-   \eta  $ & $ -\frac{\sin(\theta_C) \left(a_1+6 a_3+2 a_5-8 a_{10}-2 a_{11}+12 a_{13}-2 a_{14}+3 a_{15}-2 a_{16}-5 b_1-3 b_2-6 b_3+2 b_5+2 b_6-2 b_{11}+6 b_{14}-2 b_{16}\right)}{\sqrt{6}}$\\\hline
  			$\Omega_{cc}^{+}\to \Xi_c^+  \pi^+   K^0   K^-  $ & $ \sin(\theta_C) \left(a_1-a_5-2 a_8+a_9-6 a_{13}-2 a_{15}+a_{16}+b_1-b_2-b_5+b_6-b_9+b_{16}\right)$\\\hline
  			$\Omega_{cc}^{+}\to \Xi_c^+  \pi^0   \pi^0   \pi^0  $ & $ \frac{3\sin(\theta_C) \left(a_1+2 a_3-2 a_{14}-a_{15}-b_1+b_2-2 b_3-2 b_{14}\right)}{\sqrt{2}}$\\\hline
  			$\Omega_{cc}^{+}\to \Xi_c^+  \pi^0   \pi^0   \eta  $ & $ -\frac{\sin(\theta_C) \left(a_1+6 a_3+2 a_5-8 a_{10}-2 a_{11}+12 a_{13}-2 a_{14}+3 a_{15}-2 a_{16}-b_1+b_2-6 b_3-2 b_5+2 b_6-2 b_{11}+6 b_{14}-2 b_{16}\right)}{ \sqrt{6}}$\\\hline
  			$\Omega_{cc}^{+}\to \Xi_c^+  \pi^0   K^+   K^-  $ & $ \frac{\sin(\theta_C) \left(2 a_3-a_5+2 a_7-2 a_8-a_9-6 a_{13}-2 a_{14}-3 a_{15}+a_{16}-2 b_1-2 b_2-2 b_3+b_5+b_6-b_9-2 b_{14}+b_{16}\right)}{\sqrt{2}}$\\\hline
  			$\Omega_{cc}^{+}\to \Xi_c^+  \pi^0   K^0   \overline K^0  $ & $ \frac{\sin(\theta_C) \left(2 a_3+a_7+2 a_8-2 a_9-a_{11}+2 a_{12}+6 a_{13}-2 a_{14}+a_{15}-a_{16}-2 b_3-b_7+b_{11}-2 b_{14}+b_{16}\right)}{\sqrt{2}}$\\\hline
  			$\Omega_{cc}^{+}\to \Xi_c^+  \pi^-   K^+   \overline K^0  $ & $ \sin(\theta_C) \left(-a_1+a_7-2 a_8+a_{11}-2 a_{12}-6 a_{13}-2 a_{15}+a_{16}-b_1+b_2+b_7-b_{11}-b_{16}\right)$\\\hline
  			$\Omega_{cc}^{+}\to \Xi_c^0  \pi^+   \pi^+   \pi^-  $ & $ 2\sin(\theta_C) \left(a_1+2 a_3-2 a_{14}-a_{15}+b_1+b_2+2 b_3-2 b_{14}\right)$\\\hline
    			$\Omega_{cc}^{+}\to \Xi_c^0  \pi^+   \pi^0   \pi^0  $ & $ \sin(\theta_C) \left(a_1+2 a_3-2 a_{14}-a_{15}+b_1+b_2+2 b_3-2 b_{14}\right)$\\\hline
  			$\Omega_{cc}^{+}\to \Xi_c^0  \pi^+   \pi^0   \eta  $ & $ -\frac{2 \left(b_1+b_2-b_5\right) \sin(\theta_C)}{\sqrt{3}}$\\\hline
  			$\Omega_{cc}^{+}\to \Xi_c^0  \pi^+   K^+   K^-  $ & $ -\sin(\theta_C) \left(a_1-2 a_3-2 a_7+2 a_9+2 a_{14}+a_{15}+b_1+b_2-2 b_3-2 b_7+2 b_9+2 b_{14}\right)$\\\hline
  			$\Omega_{cc}^{+}\to \Xi_c^0  \pi^0   K^+   \overline K^0  $ & $ \frac{\sin(\theta_C) \left(2 a_1-a_5-a_7+a_9-a_{11}+2 a_{12}-2 b_2+b_5+b_6-b_7+b_9-b_{11}+2 b_{16}\right)}{\sqrt{2}}$\\\hline
  			  		\end{tabular}
  		\end{tiny}
\end{table}

\begin{table}
\begin{tiny}
\caption{Doubly Cabibbo-suppressed channels for doubly charmed baryon decays into an antitriplet charmed baryon and three light mesons.}\label{tab:First_type_doubly}
\begin{tabular}{|c|c|c|c|c|c|c|c}\hline\hline
  			channel & amplitude \\\hline
  			\hline
  			$\Xi_{cc}^{++}\to \Lambda_c^+  \pi^+   \pi^0   K^0  $ & $ -\sqrt{2} \left(b_2-b_6+b_9\right) \sin(\theta_C)^2$\\\hline
  			$\Xi_{cc}^{++}\to \Lambda_c^+  \pi^+   \pi^-   K^+  $ & $ \sin(\theta_C)^2 \left(a_1+2 a_3-a_5+a_7-2 a_8-4 a_{10}-2 a_{11}+2 a_{12}+b_1+b_2+2 b_3-b_5-b_6+b_7\right)$\\\hline
  			$\Xi_{cc}^{++}\to \Lambda_c^+  \pi^+   K^0   \eta  $ & $ -\sqrt{\frac{2}{3}} \sin(\theta_C)^2 \left(a_1-a_7+a_9-b_1-b_6+b_7\right)$\\\hline
  			$\Xi_{cc}^{++}\to \Lambda_c^+  \pi^0   \pi^0   K^+  $ & $ \sin(\theta_C)^2 \left(a_1+2 a_3-a_5+a_7-2 a_8-4 a_{10}-2 a_{11}+2 a_{12}+b_1-b_2+2 b_3-b_5+b_6+b_7-2 b_9\right)$\\\hline
  			$\Xi_{cc}^{++}\to \Lambda_c^+  \pi^0   K^+   \eta  $ & $ -\frac{\sin(\theta_C)^2 \left(a_1-a_7+a_9+b_1-b_2-b_7-b_9+2 b_{11}\right)}{\sqrt{3}}$\\\hline
  			$\Xi_{cc}^{++}\to \Xi_c^+  \pi^+   K^0   K^0  $ & $ 2\sin(\theta_C)^2 \left(-a_1+a_7-a_9+b_1+b_2-b_7+b_9\right)$\\\hline
  			$\Xi_{cc}^{++}\to \Xi_c^+  \pi^0   K^+   K^0  $ & $ -\frac{\sin(\theta_C)^2 \left(2 a_1-a_5-a_7+a_9-a_{11}+2 a_{12}-b_5-b_6+b_7-b_9+b_{11}\right)}{\sqrt{2}}$\\\hline
  			$\Xi_{cc}^{++}\to \Xi_c^+  \pi^-   K^+   K^+  $ & $ 2\sin(\theta_C)^2 \left(a_1-a_5-a_{11}+2 a_{12}+b_1+b_2-b_5-b_6+b_{11}\right)$\\\hline
  			$\Xi_{cc}^{++}\to \Xi_c^0  \pi^+   K^+   K^0  $ & $ \sin(\theta_C)^2 \left(a_5-a_7+a_9+a_{11}-2 a_{12}+b_5-b_6-b_7+b_9+b_{11}\right)$\\\hline
  			$\Xi_{cc}^{++}\to \Xi_c^0  \pi^0   K^+   K^+  $ & $ \sqrt{2}\sin(\theta_C)^2 \left(a_5-a_7+a_9+a_{11}-2 a_{12}+b_5-b_6-b_7+b_9+b_{11}\right)$\\\hline
  			$\Xi_{cc}^{+}\to \Lambda_c^+  \pi^+   \pi^-   K^0  $ & $ \sin(\theta_C)^2 \left(a_1+2 a_3-a_5+a_7-2 a_8-4 a_{10}-2 a_{11}+2 a_{12}-b_1-b_2-2 b_3+b_5+b_6-b_7\right)$\\\hline
  			$\Xi_{cc}^{+}\to \Lambda_c^+  \pi^0   \pi^0   K^0  $ & $ \sin(\theta_C)^2 \left(a_1+2 a_3-a_5+a_7-2 a_8-4 a_{10}-2 a_{11}+2 a_{12}-b_1+b_2-2 b_3+b_5-b_6-b_7+2 b_9\right)$\\\hline
  			$\Xi_{cc}^{+}\to \Lambda_c^+  \pi^0   \pi^-   K^+  $ & $ -\sqrt{2} \left(b_2-b_6+b_9\right) \sin(\theta_C)^2$\\\hline
  			$\Xi_{cc}^{+}\to \Lambda_c^+  \pi^0   K^0   \eta  $ & $ \frac{\sin(\theta_C)^2 \left(a_1-a_7+a_9-b_1+b_2+b_7+b_9-2 b_{11}\right)}{\sqrt{3}}$\\\hline
  			$\Xi_{cc}^{+}\to \Lambda_c^+  \pi^-   K^+   \eta  $ & $ -\sqrt{\frac{2}{3}} \sin(\theta_C)^2 \left(a_1-a_7+a_9+b_1+b_6-b_7\right)$\\\hline
  			$\Xi_{cc}^{+}\to \Xi_c^+  \pi^0   K^0   K^0  $ & $ \sqrt{2}\sin(\theta_C)^2 \left(a_5-a_7+a_9+a_{11}-2 a_{12}-b_5+b_6+b_7-b_9-b_{11}\right)$\\\hline
  			$\Xi_{cc}^{+}\to \Xi_c^+  \pi^-   K^+   K^0  $ & $ \sin(\theta_C)^2 \left(-a_5+a_7-a_9-a_{11}+2 a_{12}+b_5-b_6-b_7+b_9+b_{11}\right)$\\\hline
  			$\Xi_{cc}^{+}\to \Xi_c^0  \pi^+   K^0   K^0  $ & $ 2\sin(\theta_C)^2 \left(-a_1+a_5+a_{11}-2 a_{12}+b_1+b_2-b_5-b_6+b_{11}\right)$\\\hline
  			$\Xi_{cc}^{+}\to \Xi_c^0  \pi^0   K^+   K^0  $ & $ -\frac{\sin(\theta_C)^2 \left(2 a_1-a_5-a_7+a_9-a_{11}+2 a_{12}+b_5+b_6-b_7+b_9-b_{11}\right)}{\sqrt{2}}$\\\hline
  			$\Xi_{cc}^{+}\to \Xi_c^0  \pi^-   K^+   K^+  $ & $ 2\sin(\theta_C)^2 \left(a_1-a_7+a_9+b_1+b_2-b_7+b_9\right)$\\\hline
  			$\Omega_{cc}^{+}\to \Lambda_c^+  \pi^+   \pi^0   \pi^-  $ & $ \sqrt{2} \left(-2 b_{14}+b_{16}\right) \sin(\theta_C)^2$\\\hline
  			$\Omega_{cc}^{+}\to \Lambda_c^+  \pi^+   \pi^-   \eta  $ & $ \sqrt{\frac{2}{3}} \left(4 a_{10}+2 a_{11}-2 a_{12}-6 a_{13}-2 a_{14}-3 a_{15}+a_{16}\right) \sin(\theta_C)^2$\\\hline
  			$\Omega_{cc}^{+}\to \Lambda_c^+  \pi^+   K^0   K^-  $ & $ \sin(\theta_C)^2 \left(-a_5+a_7-2 a_8-6 a_{13}-2 a_{15}+a_{16}+b_5+b_6-b_7-b_{16}\right)$\\\hline
  			$\Omega_{cc}^{+}\to \Lambda_c^+  \pi^0   \pi^0   \pi^0  $ & $ -3\sqrt{2}\left(2 b_{14}-b_{16}\right) \sin(\theta_C)^2$\\\hline
  			$\Omega_{cc}^{+}\to \Lambda_c^+  \pi^0   \pi^0   \eta  $ & $ \sqrt{\frac{2}{3}}\left(4 a_{10}+2 a_{11}-2 a_{12}-6 a_{13}-2 a_{14}-3 a_{15}+a_{16}\right) \sin(\theta_C)^2$\\\hline
  			$\Omega_{cc}^{+}\to \Lambda_c^+  \pi^0   K^+   K^-  $ & $ -\frac{\sin(\theta_C)^2 \left(a_5-a_7+2 a_8+6 a_{13}+2 a_{15}-a_{16}+b_5-b_6-b_7+2 b_9+4 b_{14}-b_{16}\right)}{\sqrt{2}}$\\\hline
  			$\Omega_{cc}^{+}\to \Lambda_c^+  \pi^0   K^0   \overline K^0  $ & $ \frac{\sin(\theta_C)^2 \left(a_5-a_7+2 a_8+6 a_{13}+2 a_{15}-a_{16}-b_5+b_6+b_7-2 b_9-4 b_{14}+b_{16}\right)}{\sqrt{2}}$\\\hline
  			$\Omega_{cc}^{+}\to \Lambda_c^+  \pi^0   \eta   \eta  $ & $ -\frac{\sqrt{2}\left(4 b_9-4 b_{11}+6 b_{14}+b_{16}\right) \sin(\theta_C)^2}{3}$\\\hline
  			$\Omega_{cc}^{+}\to \Lambda_c^+  \pi^-   K^+   \overline K^0  $ & $ \sin(\theta_C)^2 \left(-a_5+a_7-2 a_8-6 a_{13}-2 a_{15}+a_{16}-b_5-b_6+b_7+b_{16}\right)$\\\hline
  			$\Omega_{cc}^{+}\to \Lambda_c^+  K^+   K^-   \eta  $ & $ \frac{\sin(\theta_C)^2 \left(a_5-a_7+6 a_8-2 a_9+8 a_{10}+2 a_{11}+6 a_{13}-4 a_{14}-a_{16}+b_5-b_6-b_7+2 b_{11}-b_{16}\right)}{\sqrt{6}}$\\\hline
  			$\Omega_{cc}^{+}\to \Lambda_c^+  K^0   \overline K^0   \eta  $ & $ \frac{\sin(\theta_C)^2 \left(a_5-a_7+6 a_8-2 a_9+8 a_{10}+2 a_{11}+6 a_{13}-4 a_{14}-a_{16}-b_5+b_6+b_7-2 b_{11}+b_{16}\right)}{\sqrt{6}}$\\\hline
  			$\Omega_{cc}^{+}\to \Xi_c^+  \pi^+   \pi^-   K^0  $ & $ \sin(\theta_C)^2 \left(a_1+2 a_3-2 a_{14}-a_{15}-b_1-b_2-2 b_3+2 b_{14}\right)$\\\hline
  			$\Omega_{cc}^{+}\to \Xi_c^+  \pi^0   \pi^0   K^0  $ & $ \sin(\theta_C)^2 \left(a_1+2 a_3-2 a_{14}-a_{15}-b_1+b_2-2 b_3+2 b_{14}-2 b_{16}\right)$\\\hline
  			$\Omega_{cc}^{+}\to \Xi_c^+  \pi^0   \pi^-   K^+  $ & $ \sqrt{2} \left(b_{16}-b_2\right) \sin(\theta_C)^2$\\\hline
  			$\Omega_{cc}^{+}\to \Xi_c^+  \pi^0   K^0   \eta  $ & $ \frac{\sin(\theta_C)^2 \left(a_1-a_5-a_{11}+2 a_{12}-b_1+b_2+b_5-b_6+b_{11}-2 b_{16}\right)}{\sqrt{3}}$\\\hline
  			$\Omega_{cc}^{+}\to \Xi_c^+  \pi^-   K^+   \eta  $ & $ -\sqrt{\frac{2}{3}} \sin(\theta_C)^2 \left(a_1-a_5-a_{11}+2 a_{12}+b_1-b_5-b_6+b_{11}+b_{16}\right)$\\\hline
  			$\Omega_{cc}^{+}\to \Xi_c^0  \pi^+   \pi^0   K^0  $ & $ \sqrt{2} \left(b_2-b_{16}\right) \sin(\theta_C)^2$\\\hline
  			$\Omega_{cc}^{+}\to \Xi_c^0  \pi^+   \pi^-   K^+  $ & $ -\sin(\theta_C)^2 \left(a_1+2 a_3-2 a_{14}-a_{15}+b_1+b_2+2 b_3-2 b_{14}\right)$\\\hline
  			$\Omega_{cc}^{+}\to \Xi_c^0  \pi^+   K^0   \eta  $ & $ \sqrt{\frac{2}{3}} \sin(\theta_C)^2 \left(a_1-a_5-a_{11}+2 a_{12}-b_1+b_5+b_6-b_{11}-b_{16}\right)$\\\hline
  			$\Omega_{cc}^{+}\to \Xi_c^0  \pi^0   \pi^0   K^+  $ & $ -\sin(\theta_C)^2 \left(a_1+2 a_3-2 a_{14}-a_{15}+b_1-b_2+2 b_3-2 b_{14}+2 b_{16}\right)$\\\hline
  			$\Omega_{cc}^{+}\to \Xi_c^0  \pi^0   K^+   \eta  $ & $ \frac{\sin(\theta_C)^2 \left(a_1-a_5-a_{11}+2 a_{12}+b_1-b_2-b_5+b_6-b_{11}+2 b_{16}\right)}{\sqrt{3}}$\\\hline
  			\hline
\end{tabular}
\end{tiny}
\end{table}

From these amplitudes, we can find the relations for decay widths in the SU(3) symmetry limit:
\begin{small}
\begin{eqnarray}
	&&\Gamma(\Omega_{cc}^{+}\to\Lambda_c^+\pi^+ \overline K^0 K^- )= \Gamma(\Omega_{cc}^{+}\to\Lambda_c^+\pi^0 \overline K^0 \overline K^0 ) = 3\Gamma(\Omega_{cc}^{+}\to\Lambda_c^+\overline K^0 \overline K^0 \eta ) \,,\nn\\
    &&\Gamma(\Xi_{cc}^{++}\to\Lambda_c^+\pi^+ \pi^+ K^- )=4\Gamma(\Xi_{cc}^{+}\to\Lambda_c^+\pi^+ \pi^0 K^- )= 4\Gamma(\Xi_{cc}^{++}\to\Lambda_c^+\pi^+ \pi^0 \overline K^0 ) \,,\nn\\
    &&\Gamma(\Xi_{cc}^{++}\to\Xi_c^+\pi^+ \pi^0 \eta )= { }\Gamma(\Xi_{cc}^{+}\to\Xi_c^0\pi^+ \pi^0 \eta ) = \frac{1}{4}\Gamma(\Xi_{cc}^{++}\to\Xi_c^0\pi^+ \pi^+ \eta ) = \frac{1}{3}\Gamma(\Xi_{cc}^{++}\to\Xi_c^0\pi^+ K^+ \overline K^0 ) \,,\nn\\
    &&\Gamma(\Omega_{cc}^{+}\to\Xi_c^+\pi^+ \pi^0 K^- )= { }\Gamma(\Omega_{cc}^{+}\to\Xi_c^0\pi^+ \pi^0 \overline K^0 )=\Gamma(\Omega_{cc}^{+}\to\Xi_c^+\pi^0 \pi^- K^+ ) = \frac{1}{4}\Gamma(\Omega_{cc}^{+}\to\Xi_c^0\pi^+ \pi^+ K^- ) \,,\nn\\
    && { }\Gamma(\Omega_{cc}^{+}\to\Xi_c^0\pi^+ \pi^+ \pi^- )= 4\Gamma(\Omega_{cc}^{+}\to\Xi_c^0\pi^+ \pi^0 \pi^0 ) \,,\quad
    \Gamma(\Xi_{cc}^{++}\to\Lambda_c^+\pi^+ \pi^0 K^0 )= { }\Gamma(\Xi_{cc}^{+}\to\Lambda_c^+\pi^0 \pi^- K^+ ) \,,\nn\\
    &&\Gamma(\Xi_{cc}^{+}\to\Xi_c^0\pi^+ \pi^+ \pi^- )= 4\Gamma(\Xi_{cc}^{+}\to\Xi_c^0\pi^+ \pi^0 \pi^0 ) \,,\quad
    \Gamma(\Xi_{cc}^{+}\to\Xi_c^0\pi^+ \pi^- K^+ )= { }\Gamma(\Omega_{cc}^{+}\to\Xi_c^0\pi^+ K^+ K^- ) \,,\nn\\
    &&\Gamma(\Xi_{cc}^{+}\to\Xi_c^+\pi^+ \pi^0 \pi^- )= \frac{2}{3}\Gamma(\Xi_{cc}^{+}\to\Xi_c^+\pi^0 \pi^0 \pi^0 ) \,,\quad
    \Gamma(\Xi_{cc}^{+}\to\Xi_c^+\pi^+ \pi^- K^0 )= { }\Gamma(\Omega_{cc}^{+}\to\Lambda_c^+K^+ \overline K^0 K^- ) \,,\nn\\
    &&\Gamma(\Xi_{cc}^{++}\to\Xi_c^+\pi^+ \pi^+ \pi^- )= 4\Gamma(\Xi_{cc}^{++}\to\Xi_c^+\pi^+ \pi^0 \pi^0 ) \,,\quad
    \Gamma(\Xi_{cc}^{+}\to\Lambda_c^+\pi^+ \pi^0 \pi^- )= \frac{2}{3}\Gamma(\Xi_{cc}^{+}\to\Lambda_c^+\pi^0 \pi^0 \pi^0 ) \,,\nn\\
    &&\Gamma(\Xi_{cc}^{+}\to\Xi_c^+\pi^0 K^0 K^0 )= \Gamma(\Xi_{cc}^{+}\to\Xi_c^+\pi^- K^+ K^0 ) \,,\quad
    \Gamma(\Xi_{cc}^{++}\to\Xi_c^0\pi^+ K^+ K^0 )= \Gamma(\Xi_{cc}^{++}\to\Xi_c^0\pi^0 K^+ K^+ ) \,,\nn\\
    &&\Gamma(\Xi_{cc}^{+}\to\Lambda_c^+\pi^+ K^0 K^- )= { }\Gamma(\Omega_{cc}^{+}\to\Xi_c^+\pi^- K^+ \overline K^0 ) \,,\quad
    \Gamma(\Xi_{cc}^{+}\to\Lambda_c^+\pi^- K^+ \overline K^0 )= { }\Gamma(\Omega_{cc}^{+}\to\Xi_c^+\pi^+ K^0 K^- ) \,,\nn\\
    &&\Gamma(\Xi_{cc}^{++}\to\Lambda_c^+\pi^+ K^+ K^- )= { }\Gamma(\Xi_{cc}^{++}\to\Xi_c^+\pi^+ \pi^- K^+ ) \,,\quad
    \Gamma(\Omega_{cc}^{+}\to\Lambda_c^+\pi^+ \pi^0 \pi^- )= \frac{2}{3}\Gamma(\Omega_{cc}^{+}\to\Lambda_c^+\pi^0 \pi^0 \pi^0 )\,,\nn\\
    &&\Gamma(\Xi_{cc}^{++}\to\Xi_c^0\pi^+ \pi^+ K^0 )=  4\Gamma(\Xi_{cc}^{++}\to\Xi_c^0\pi^+ \pi^0 K^+ ) \,,\quad
    \Gamma(\Xi_{cc}^{++}\to\Lambda_c^+\pi^+ \pi^+ \pi^- )= 4\Gamma(\Xi_{cc}^{++}\to\Lambda_c^+\pi^+ \pi^0 \pi^0 )\,,\nn\\
    &&\Gamma(\Omega_{cc}^{+}\to\Lambda_c^+\pi^+ \pi^- \eta )= 2\Gamma(\Omega_{cc}^{+}\to\Lambda_c^+\pi^0 \pi^0 \eta ) \,,\quad
    \Gamma(\Omega_{cc}^{+}\to\Xi_c^+\pi^+ \pi^0 \pi^- )= \frac{2}{3}\Gamma(\Omega_{cc}^{+}\to\Xi_c^+\pi^0 \pi^0 \pi^0 ) \,.
\end{eqnarray}
\end{small}

Several decay channels such like $\Xi_{cc}^{+}\to\Lambda_c^+\pi^+ \pi^0 K^-$, $\Omega_{cc}^{+}\to\Lambda_c^+\pi^+ \overline K^0 K^-$, and $\Xi_{cc}^{++}\to\Lambda_c^+\pi^+ \pi^0 \overline K^0$ might be helpful to search for $\Xi_{cc}^+$, $\Omega_{cc}^+$, and new decay channels for $\Xi_{cc}^{++}$ at LHC, since their branching fractions are sizeable, and the productions are easy to be identified.

It should be noted that the above relationships between decay widths are obtained in the flavor SU(3) symmetry limit, in which the mass differences between final state hadrons have been ignored. We have removed the channels kinematically prohibited. Although the influence of identical particles on phase space integration has been considered, these relationships will be modified when calculating the kinematic corrections rigorously. We can also investigate the light vector mesons in the final states except for the light pseudo-scalar mesons since they have the same quark constitutions with different quantum numbers. The effective Hamitonian is the same as the case of light vector mesons, so one can easily get the amplitudes for a singly charmed baryon and three light vector mesons by replacing the final state mesons $\pi \rightarrow \rho, K \rightarrow K^{*}$, and $\eta \rightarrow \omega$.

\subsection{Decays into an octet light baryon with a charmed meson and two light pseudo-scalar mesons}
\begin{figure}
\includegraphics[width=1\columnwidth]{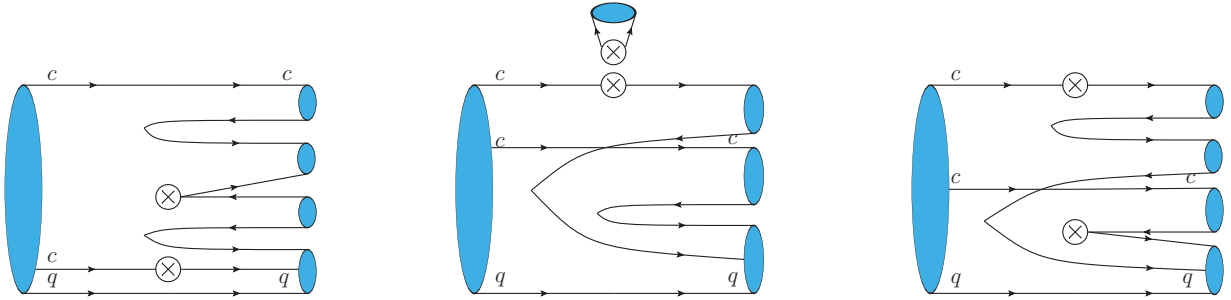}
\caption{ Feynman diagrams for $\Xi_{cc}$ and $\Omega_{cc}$ decays into an octet light baryon with a charmed meson and two light pseudo-scalar mesons. }
\label{feynman2}
\end{figure}

The effective Hamitonian for $\Xi_{cc}$ and $\Omega_{cc}$ decay into an octet light baryon with a charmed meson and two light pseudo-scalar mesons can be shown as
\begin{footnotesize}
\begin{eqnarray}\label{H:2}
 &&{\cal H}_{\textit{eff}}= \nn\\
   &&+c_1  (T_{cc})^i  \epsilon_{ijk} (T_8)^{k}_{l}  \overline D^j   M^{l}_{m}  M^{r}_{n}  (H_{\overline6})^{mn}_{r}
     +c_2  (T_{cc})^i  \epsilon_{ijk} (T_8)^{k}_{l}  \overline D^j   M^{n}_{r}  M^{r}_{m}  (H_{\overline6})^{lm}_{n}
     +c_3  (T_{cc})^i  \epsilon_{ijk} (T_8)^{k}_{l}  \overline D^l   M^{j}_{m}  M^{r}_{n}  (H_{\overline6})^{mn}_{r} \nn\\
   &&+c_4  (T_{cc})^i  \epsilon_{ijk} (T_8)^{k}_{l}  \overline D^m   M^{j}_{n}  M^{l}_{r}  (H_{\overline6})^{nr}_{m}
     +c_5  (T_{cc})^i  \epsilon_{ijk} (T_8)^{k}_{l}  \overline D^n   M^{j}_{m}  M^{r}_{n}  (H_{\overline6})^{lm}_{r}
     +c_6  (T_{cc})^i  \epsilon_{ijk} (T_8)^{k}_{l}  \overline D^l   M^{m}_{n}  M^{n}_{r}  (H_{\overline6})^{jr}_{m} \nn\\
   &&+c_7  (T_{cc})^i  \epsilon_{ijk} (T_8)^{k}_{l}  \overline D^l   M^{m}_{n}  M^{r}_{m}  (H_{\overline6})^{jn}_{r}
     +c_8  (T_{cc})^i  \epsilon_{ijk} (T_8)^{k}_{l}  \overline D^m   M^{l}_{m}  M^{n}_{r}  (H_{\overline6})^{jr}_{n}
     +c_9  (T_{cc})^i  \epsilon_{ijk} (T_8)^{k}_{l}  \overline D^n   M^{l}_{m}  M^{m}_{r}  (H_{\overline6})^{jr}_{n} \nn\\
   &&+c_{10}  (T_{cc})^i  \epsilon_{ijk} (T_8)^{k}_{l}  \overline D^r   M^{l}_{m}  M^{n}_{r}  (H_{\overline6})^{jm}_{n}
     +c_{11}  (T_{cc})^i  \epsilon_{ijk} (T_8)^{k}_{l}  \overline D^m   M^{n}_{r}  M^{r}_{n}  (H_{\overline6})^{jl}_{m}
     +c_{12}  (T_{cc})^i  \epsilon_{ijk} (T_8)^{k}_{l}  \overline D^n   M^{m}_{r}  M^{r}_{n}  (H_{\overline6})^{jl}_{m} \nn\\
   &&+c_{13}  (T_{cc})^l  \epsilon_{ijk} (T_8)^{k}_{l}  \overline D^i   M^{j}_{m}  M^{r}_{n}  (H_{\overline6})^{mn}_{r}
     +c_{14}  (T_{cc})^r  \epsilon_{ijk} (T_8)^{k}_{l}  \overline D^i   M^{j}_{m}  M^{l}_{n}  (H_{\overline6})^{mn}_{r}
     +c_{15}  (T_{cc})^n  \epsilon_{ijk} (T_8)^{k}_{l}  \overline D^i   M^{j}_{r}  M^{r}_{m}  (H_{\overline6})^{lm}_{n} \nn\\
   &&+c_{16}  (T_{cc})^r  \epsilon_{ijk} (T_8)^{k}_{l}  \overline D^i   M^{j}_{r}  M^{n}_{m}  (H_{\overline6})^{lm}_{n}
     +c_{17}  (T_{cc})^r  \epsilon_{ijk} (T_8)^{k}_{l}  \overline D^i   M^{j}_{m}  M^{n}_{r}  (H_{\overline6})^{lm}_{n}
     +c_{18}  (T_{cc})^l  \epsilon_{ijk} (T_8)^{k}_{l}  \overline D^i   M^{r}_{m}  M^{n}_{r}  (H_{\overline6})^{jm}_{n} \nn\\
   &&+c_{19}  (T_{cc})^n  \epsilon_{ijk} (T_8)^{k}_{l}  \overline D^i   M^{l}_{r}  M^{r}_{m}  (H_{\overline6})^{jm}_{n}
     +c_{20}  (T_{cc})^r  \epsilon_{ijk} (T_8)^{k}_{l}  \overline D^i   M^{l}_{r}  M^{n}_{m}  (H_{\overline6})^{jm}_{n}
     +c_{21}  (T_{cc})^r  \epsilon_{ijk} (T_8)^{k}_{l}  \overline D^i   M^{l}_{m}  M^{n}_{r}  (H_{\overline6})^{jm}_{n} \nn\\
   &&+c_{22}  (T_{cc})^m  \epsilon_{ijk} (T_8)^{k}_{l}  \overline D^i   M^{n}_{r}  M^{r}_{n}  (H_{\overline6})^{jl}_{m}
     +c_{23}  (T_{cc})^r  \epsilon_{ijk} (T_8)^{k}_{l}  \overline D^i   M^{n}_{r}  M^{m}_{n}  (H_{\overline6})^{jl}_{m}
     +c_{24}  (T_{cc})^l  \epsilon_{ijk} (T_8)^{k}_{l}  \overline D^r   M^{i}_{m}  M^{j}_{n}  (H_{\overline6})^{mn}_{r} \nn\\
   &&+c_{25}  (T_{cc})^r  \epsilon_{ijk} (T_8)^{k}_{l}  \overline D^l   M^{i}_{m}  M^{j}_{n}  (H_{\overline6})^{mn}_{r}
     +c_{26}  (T_{cc})^n  \epsilon_{ijk} (T_8)^{k}_{l}  \overline D^r   M^{i}_{r}  M^{j}_{m}  (H_{\overline6})^{lm}_{n}
     +c_{27}  (T_{cc})^r  \epsilon_{ijk} (T_8)^{k}_{l}  \overline D^n   M^{i}_{r}  M^{j}_{m}  (H_{\overline6})^{lm}_{n} \nn\\
   &&+c_{28}  (T_{cc})^l  \epsilon_{ijk} (T_8)^{k}_{l}  \overline D^n   M^{i}_{r}  M^{r}_{m}  (H_{\overline6})^{jm}_{n}
     +c_{29}  (T_{cc})^l  \epsilon_{ijk} (T_8)^{k}_{l}  \overline D^r   M^{i}_{r}  M^{n}_{m}  (H_{\overline6})^{jm}_{n}
     +c_{30}  (T_{cc})^l  \epsilon_{ijk} (T_8)^{k}_{l}  \overline D^r   M^{i}_{m}  M^{n}_{r}  (H_{\overline6})^{jm}_{n} \nn\\
   &&+c_{31}  (T_{cc})^n  \epsilon_{ijk} (T_8)^{k}_{l}  \overline D^l   M^{i}_{r}  M^{r}_{m}  (H_{\overline6})^{jm}_{n}
     +c_{32}  (T_{cc})^r  \epsilon_{ijk} (T_8)^{k}_{l}  \overline D^l   M^{i}_{r}  M^{n}_{m}  (H_{\overline6})^{jm}_{n}
     +c_{33}  (T_{cc})^r  \epsilon_{ijk} (T_8)^{k}_{l}  \overline D^l   M^{i}_{m}  M^{n}_{r}  (H_{\overline6})^{jm}_{n} \nn\\
   &&+c_{34}  (T_{cc})^n  \epsilon_{ijk} (T_8)^{k}_{l}  \overline D^r   M^{i}_{r}  M^{l}_{m}  (H_{\overline6})^{jm}_{n}
     +c_{35}  (T_{cc})^n  \epsilon_{ijk} (T_8)^{k}_{l}  \overline D^r   M^{i}_{m}  M^{l}_{r}  (H_{\overline6})^{jm}_{n}
     +c_{36}  (T_{cc})^r  \epsilon_{ijk} (T_8)^{k}_{l}  \overline D^n   M^{i}_{m}  M^{l}_{r}  (H_{\overline6})^{jm}_{n} \nn\\
   &&+c_{37}  (T_{cc})^r  \epsilon_{ijk} (T_8)^{k}_{l}  \overline D^n   M^{i}_{r}  M^{l}_{m}  (H_{\overline6})^{jm}_{n}
     +c_{38}  (T_{cc})^m  \epsilon_{ijk} (T_8)^{k}_{l}  \overline D^n   M^{i}_{m}  M^{r}_{n}  (H_{\overline6})^{jl}_{r}
     +c_{39}  (T_{cc})^m  \epsilon_{ijk} (T_8)^{k}_{l}  \overline D^n   M^{i}_{n}  M^{r}_{m}  (H_{\overline6})^{jl}_{r} \nn\\
   &&+c_{40}  (T_{cc})^m  \epsilon_{ijk} (T_8)^{k}_{l}  \overline D^n   M^{i}_{r}  M^{r}_{m}  (H_{\overline6})^{jl}_{n}
     +c_{41}  (T_{cc})^m  \epsilon_{ijk} (T_8)^{k}_{l}  \overline D^n   M^{i}_{r}  M^{r}_{n}  (H_{\overline6})^{jl}_{m}
     +c_{42}  (T_{cc})^l  \epsilon_{ijk} (T_8)^{k}_{l}  \overline D^m   M^{n}_{m}  M^{r}_{n}  (H_{\overline6})^{ij}_{r} \nn\\
   &&+c_{43}  (T_{cc})^l  \epsilon_{ijk} (T_8)^{k}_{l}  \overline D^m   M^{n}_{r}  M^{r}_{n}  (H_{\overline6})^{ij}_{m}
     +c_{44}  (T_{cc})^m  \epsilon_{ijk} (T_8)^{k}_{l}  \overline D^l   M^{n}_{m}  M^{r}_{n}  (H_{\overline6})^{ij}_{r}
     +c_{45}  (T_{cc})^m  \epsilon_{ijk} (T_8)^{k}_{l}  \overline D^l   M^{r}_{n}  M^{n}_{r}  (H_{\overline6})^{ij}_{m} \nn\\
   &&+c_{46}  (T_{cc})^m  \epsilon_{ijk} (T_8)^{k}_{l}  \overline D^n   M^{l}_{m}  M^{r}_{n}  (H_{\overline6})^{ij}_{r}
     +c_{47}  (T_{cc})^m  \epsilon_{ijk} (T_8)^{k}_{l}  \overline D^n   M^{l}_{r}  M^{r}_{m}  (H_{\overline6})^{ij}_{n}
     +c_{48}  (T_{cc})^m  \epsilon_{ijk} (T_8)^{k}_{l}  \overline D^n   M^{l}_{n}  M^{r}_{m}  (H_{\overline6})^{ij}_{r} \nn\\
   &&+\Big( (H_{\overline6}) \to (H_{15}), c\to d\Big) \,.
\end{eqnarray}
\end{footnotesize}
The last line represents the implicit terms corresponding to $(H_{15})$ by simply replacing $(H_{\overline6}) \to (H_{15})$ and $c\to d$. Here the $c_i$ and $d_i$ are SU(3) irreducible nonperturbative amplitudes. It is obvious that for the nonleptonic four-body decay of doubly charmed baryon, much more terms have shown up in the effective Hamiltonian compared with nonleptonic three-body decay~\cite{Shi:2017dto}.  Using symmetry properties of indices in Eq.(\ref{H:2}), one can reduce the number of irreducible nonperturbative amplitudes. Feynman diagrams for some channels are given in Fig.~\ref{feynman2}.

Since there are so many channels and the expressions for their amplitudes are desperately large, we do not show their explicit formulas here. By expanding this effective Hamitonian, several relations for decay widths can be found in the SU(3) symmetry limit:
\begin{eqnarray}
    &&\Gamma(\Xi_{cc}^{++}\to D^0 \Lambda^0 \pi^+  \pi^+   )= 4\Gamma(\Xi_{cc}^{++}\to D^+ \Lambda^0 \pi^+  \pi^0  )= 4\Gamma(\Xi_{cc}^{+}\to D^0 \Lambda^0 \pi^+  \pi^0 ) \,,\nn\\
    &&\Gamma(\Xi_{cc}^{++}\to D^0 \Sigma^0 \pi^+  \pi^+ )= 2\Gamma(\Xi_{cc}^{++}\to D^0 \Sigma^+ \pi^+  \pi^0 ) \,.
\end{eqnarray}
The number of relations decreases significantly due to so many terms rasing in Eq.(\ref{H:2}). The at least two orders of magnitude difference between the branching fractions of $\Xi_{c c}^{++} \rightarrow D^{+} p K^{-} \pi^{+}$ and $\Xi_{c c}^{++} \rightarrow \Lambda_{c}^{+} K^{-} \pi^{+} \pi^{+}$ requires a better theoretical understanding of the resonant and nonresonant contributions to the two $\Xi_{cc}^{++}$ decay modes as well as the decay modes shown above.


\section{Conclusions}
\label{sec:conclusions}
Up to date, the quark models have made great achievements on describing the hadron spectrum. Since the charm and bottom quarks are much heavier than the lighter ones, the dynamics inside light hadrons, singly heavy hadrons, and doubly heavy hadrons could be very different. The observation of $\Xi_{cc}^{++}$ by the LHCb Collaboration opens a new field of research studying the nature of baryons containing two heavy quarks.

In this paper, we have analyzed nonleptonic four-body decays of doubly charmed baryons under the flavor SU(3) symmetry. Two kinds of weakly decays modes which are generally interested have been considered: $\Xi_{cc}^{++}, \Xi_{cc}^+$ or $\Omega_{cc}^+$ decays into a singly charmed baryon plus three light pseudo-scalar mesons or decays into an octet light baryon with a charmed meson and two light pseudo-scalar mesons. We have found a number of relations or sum rules between decay widths, which can be examined in future measurements at experimental facilities like LHC, Belle II, and CEPC. At first sight, the tables of amplitudes and expressions of decay relations are desperately large, but once a few decay branching fractions were measured in future, these relations can provide abundant clues for the exploration of other decay modes. We hope this analysis together with experimental measurements in future would help to establish a QCD-based approach to handle the decays of doubly heavy baryons.

\section*{Acknowledgements}
The authors are grateful to Professor Wei Wang for inspiring discussions and valuable comments. This work is supported in part by National Natural Science Foundation of China under Grant nos. 12047545, and National Science Foundation of China under contract nos. 12005294.

\end{document}